# DEPOLARIZATION CURRENTS IN FRESH AND AGED CORONA POLED P(VDF-TFE) FILMS

S. N. Fedosov, A. E. Sergeeva, A. F. Butenko

*Department of Physics, Odessa National Academy of Food Technologies, Odessa, Ukraine*

*Using short circuit and open circuit modifications of the thermally stimulated depolarization current technique, relaxation currents have been measured in corona poled P(VDF-TFE) copolymers over the temperature range of 10–120 ºC in samples stored after poling for either 1 day, or 16 month. Two well structured peaks observed in aged samples at 23 and 90 ºC were attributed to relaxation of electret and ferroelectric components of the remnant polarization. In fresh samples the two components were mixed forming a broad peak at 71 ºC. Relaxation of the space charge caused inversion of the current in the open circuit mode both in fresh and aged samples. The space charge peaks have been extracted from the total current by appropriate calculations. High stability of the trapped charges has been confirmed. It was suggested that both the electret and the ferroelectric components of the remnant polarization were accompanied by space or surface charges.*

The thermally stimulated depolarization current (TSDC) method is known to be a powerful tool for studying relaxation processes in polymer electrets [1] Although the theory of the TSD currents has been developed mainly for the thermally frozen dipole polarization, it has been widely used to study molecular mobility and relaxation processes in ferroelectric polymers like PVDF and its copolymers [2-6]. Two TSDC peaks in these materials are of the main importance. One related to the glass transition in amorphous phase is observed at about −45 °C [5,6]. The origin of another peak seen in the range of 50-80 °C is not definitely known, although it is clear that several processes can contribute to its formation, such as reorientation of aligned dipoles in amorphous phase, relaxation of the ferroelectric polarization, migration of the space charge, interfacial and near-to-electrode processes [4].

Ferroelectric polymers, apart from their ferroelectricity caused by spontaneous switchable polarization in crystalline phase, posses properties of conventional polar electrets. Therefore, existence of two components of the remnant polarization can be expected, one related to ferroelectricity in crystalline phase, and another to aligned dipoles in amorphous phase, although there is no direct experimental evidence to show this. Besides, injected real charge has been shown to be important in ferroelectric polymers [7-9].

Thus, it is expected that three depolarization currents are superimposed in observed TSDC peaks in ferroelectric polymers, two caused by relaxation of electret and ferroelectric components of the remnant polarization and one related to the space charge. Here we distinguish between these three processes by analyzing short-circuit and open-circuit TSD currents in corona poled films of the copolymer of vinylidene fluoride and tetrafluoroethylene. P(VDF-TFE) has been chosen as a typical ferroelectric polymer, but the least studied one.

The study was performed on extruded and uniaxially stretched P(VDF-TFE) copolymer films of 20 μm thickness composed of 95 mol % VDF and 5 mol % TFE. The crystalline part of the polymer contained about 90% of the ferroelectric *b* phase according to the infrared spectroscopy data. All films were electroded from one side by evaporation of aluminum in vacuum and poled in a corona triode [7, 8] at a constant grid voltage of -4 kV. The samples were cooled down from 85 ºC to room temperature without removal of the field. Poled samples were subjected to a linear temperature ramp of rate 4 K min$^{-1}$ either in the short-circuit (SC), or in the open-circuit (OC) mode [1]. In the SC mode samples were short-circuited between two electrodes, while in the OC mode a 25 μm-thick FEP-Teflon spacer was used as a dielectric gap between free surface of the sample and one of the electrodes. The storage time between poling and measurement was either one day, or 16 months. The samples are here referred to as "fresh" and "aged" ones correspondingly.

Similarly to results reported by other workers on PVDF and P(VDF-TrTE) copolymers, [3–6] we observed one broad peak in the SC mode on fresh P(VDF-TFE) samples, as shown in Fig. 1. Direction of the current in this peak corresponds to relaxation of the remnant polarization. Since crystallinity of P(VDF-TFE) is about 50% and most of the molecular dipoles in the crystalline region are in the ferroelectric β phase, contributions of the electret and the ferroelectric components to formation of this peak in fresh samples seem to be comparable.

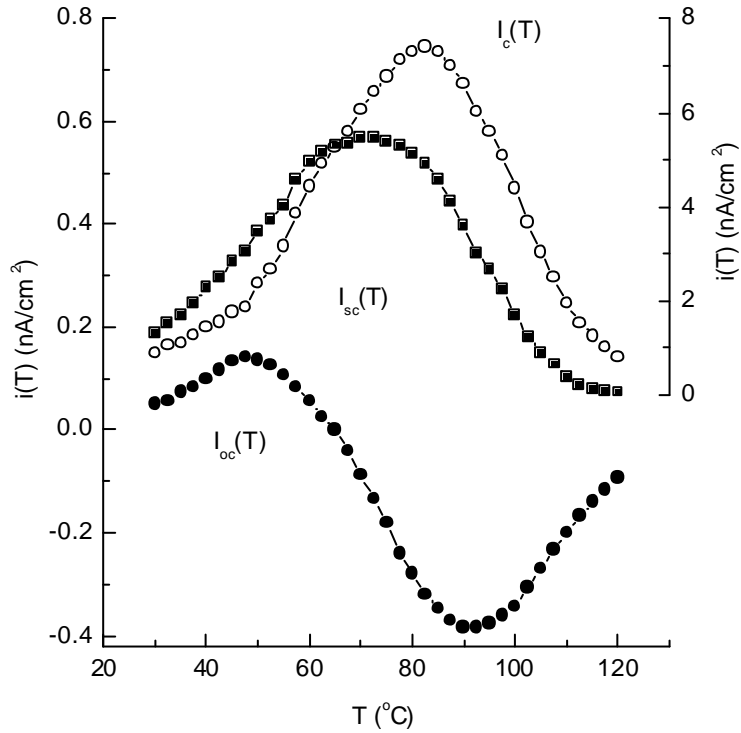

Fig. 1. TSD currents $I_{sc}(T)$ and $I_{oc}(T)$ measured on fresh poled samples in short-circuit and open circuit configutations. $I_c(T)$ is the space charge current calculated according to Eq. (1).

As for the space charge component, it is known that it either not gives any contribution to the measured TSD current in the SC mode, or its direction coincides with that of the depolarization component [1] The depolarization current in the OC mode remains invariable, while the space charge current changes its direction. Therefore, two peaks in the OC mode shown in Fig. 1 can be explained as produced by two partly overlapping and oppositely directed currents resulted from relaxation of polarization and space charge.

In order to separate the depolarization current $I_p(T)$ from the space charge one $I_c(T)$, it is reasonable to assume that polarization is uniform in the thickness direction. Since compensating charges trapped near the surfaces do not produce any current in the SC mode [1] $I_{sc}(T) = I_p(T)$ where $I_{sc}(T)$ is the experimentally measured TSD current in the SC mode. $I_c(T)$ can be calculated from experimental curves $I_{sc}(T)$ and $I_{oc}(T)$ shown in Fig. 1

$$I_c(T) = \left(1 + \frac{e_1 x_2}{e_2 x_1}\right) I_{oc}(T) - I_{sc}(T) \qquad (1)$$

where $e_1$, $x_1$, $e_2$ and $x_2$ are the dielectric constant and the thickness of the sample and the dielectric gap correspondingly. In our calculations we used $e_1=12$, $e_2=2.1$, $x_1=20$ μm, $x_2=25$ μm. It is remarkable that the $I_c(T)$ peak is positioned at higher temperature than the depolarization peak, indicating that the trapped charge is more stable than the residual polarization.

Comparing TSD currents obtained on fresh and aged poled samples, we observed a new phenomenon, namely, one broad TSD peak in the SC mode has been parted in two narrow and well separated peaks, while two pairs of oppositely directed peaks have appeared in the aged samples, instead of only one pair typical for the fresh samples, as it is shown it Fig. 2. This feature, most probably, is common for all ferroelectric polymers and does not depend on poling conditions, because we obtained similar results on P(VDF-TFE) and PVDF poled through a lime glass at 7 kV, and on samples poled by the defocused electron beam at 20 keV.

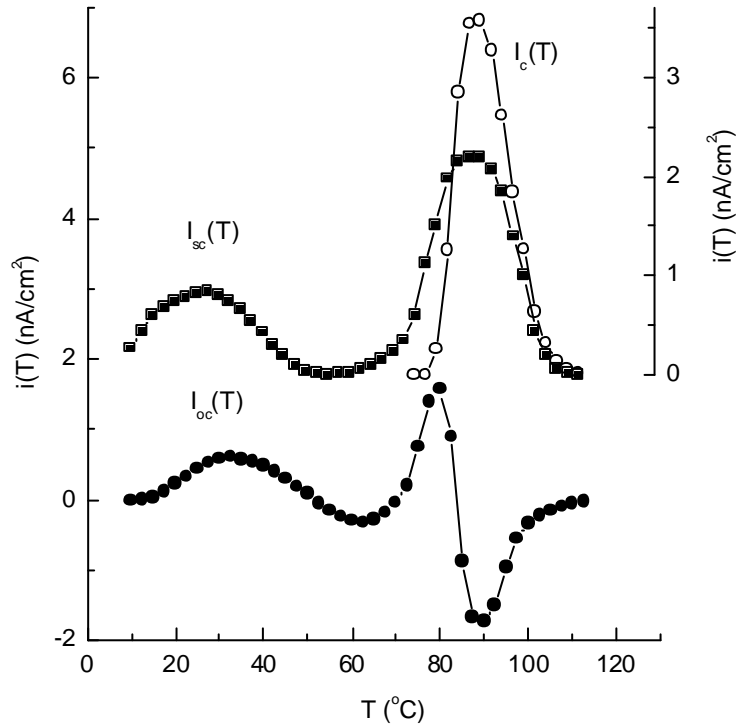

Fig. 2. TSD currents $I_{sc}(T)$ and $I_{oc}(T)$ measured in SC and OC modes on poled samples after storage for 16 month. $I_c(T)$ is the space charge current calculated according to Eq. (1).

The less thermally stable TSDC peak in the SC mode can be attributed to relaxation of the electret component of polarization, but the more stable one has obviously the ferroelectric origin. Inversion of the OC current at the high temperature region is probably caused by relaxation of the space charge. The peak of $I_c(T)$ calculated according to Eq. (1) is shown in Fig. 2 together with the measured TSD currents. It is clear from its position that the space charge is more stable than the ferroelectric polarization.

The obtained results can be explained qualitatively considering different origin of the three components of the TSD current. The electret polarization, amounting almost 50% of the remnant polarization in fresh samples, decays with time faster than the ferroelectric component. That is why two overlapping peaks in fresh samples become well separated in the aged ones, as if a slow redistribution of the remnant polarization is taking place for a long time after completion of the poling.

It is difficult to treat observed and calculated peaks quantitatively, since there is no theory of the TSD currents for ferroelectric polymers. Nevertheless, it is clear from the shape of the peaks that all three relaxation processes in P(VDF-TFE) deviate considerably from the ideal Debye case of non-interacting relaxing units. This can be taken into account by assuming that polarization decays with time according to the stretched exponential law

$$P(t) = P_o \exp\left(-\frac{t}{\bm{t}}\right)^{\bm{a}} \quad 1 \geq \bm{a} > 0 \tag{2}$$

where $\bm{t}$ is the time constant, $P_o$ is the initial polarization. If a sample is linearly heated at the rate $\bm{b}=dT/dt$, then

$$P(T) = P_o \exp\left(-\frac{1}{\bm{b}}\int_{T_o}^{T}\frac{dT'}{\bm{t}(T')}\right)^{\bm{a}} \tag{3}$$

where $T_o$ is the initial temperature. It is reasonable to assume that temperature dependence of $\bm{t}$ obeys the Arrhenius law

$$\bm{t}(T) = \bm{t}_o \exp\left(\frac{A}{kT}\right) \tag{4}$$

where $A$ is the activation energy, $k$ is Boltzmann's constant, $\bm{t}_0$ is the characteristic time. The expression for the TSD current density is obtained from Eqs. (2)-(4)

$$i(T) = -\frac{\bm{a}P_o}{\bm{t}_o}\exp\left(-\frac{A}{kT}\right)[s(T)]^{\bm{a}-1}\exp\left[-(s(T))^{\bm{a}}\right] \tag{5}$$

where

$$s(T) = \frac{1}{\bm{b}\bm{t}_o}\int_{T_o}^{T}\exp\left(-\frac{A}{kT'}\right)dT' \tag{6}$$

*Table 1*

**Parameters of relaxation processes obtained by fitting the TSDC peaks shown in Fig. 1 and Fig. 2 into Eq. (5).**

| Origin of the peak | $\bm{a}$ | $A$ (eV) | $t_0$ (s) | $T_{max}$ (°C) |
|---|---|---|---|---|
| Ferroelectric and electret polarization* | 0.24 | 2.13 | $1.1\cdot10^{-29}$ | 71.2 |
| Space charge* | 0.28 | 2.2 | $1.0\cdot10^{-29}$ | 82.8 |
| Ferroelectric polarization** | 0.52 | 2.7 | $3.4\cdot10^{-36}$ | 90.4 |
| Space charge** | 0.55 | 3.0 | $2.1\cdot10^{-40}$ | 90.5 |

\* fresh samples, ** aged samples

Results of the computer fitting of the experimentally observed and calculated TSDC peaks into Eq. (5) confirmed our conclusions on the origin and the thermal stability of relaxation processes. They show that the depolarization peak in fresh samples where electret and ferroelectric components are mixed together is a broad one ($a$=0.24), because two relaxation processes responsible for its formation are very different. Ferroelectric polarization is rather stable ($A = 2.7$ eV) and the TSDC peak caused by its relaxation is relatively narrow ($a = 0.52$). Parameters of the space charge peaks in fresh and aged samples are completely different, as if there exist two kinds of space charges, one related to the ferroelectric polarization, and another associated with the electret component. It is probable that a small peak besides the electret depolarization peak in the open-circuit mode (Fig. 2) is produced just by the electret component of the space charge. Since the glass transition temperature in P(VDF-TFE) is about -45 $^o$C, [6] alignment of dipoles in amorphous phase is not thermally frozen, as in conventional polar electrets. The preferential orientation of dipoles under these conditions can be supported by the field of the trapped charges.

In conclusion, it has been shown that two components of polarization exist in corona poled P(VDF-TFE) and probably in other ferroelectric polymers, both accompanied by corresponding space charges. The electret component, being not stable thermodynamically, decays with time until a broad TSDC peak observed in fresh poled samples is transformed in two well separated narrow peaks. The unstable electret component of the remnant polarization, can be removed by heating a poled sample to a definite temperature (about 60 $^o$C in the case of P(VDF-TFE)). It seems that the space charge, or the surface charge in the case of the uniform distribution of the residual polarization, always accompanies the dipole polarization independently of its origin.